\documentclass[12pt]{article}
\usepackage{amsmath, amssymb}
\usepackage{graphicx,psfrag}
\usepackage{enumerate}
\RequirePackage[caption=false]{subfig}
\usepackage{natbib}
\usepackage{dsfont}
\usepackage{url} 

\newcommand{\blind}{0}

\newcommand{\T}{^{\mathsf{T}}}

\DeclareMathOperator{\nach}{ne}

\DeclareMathOperator{\pa}{pa}

\DeclareMathOperator{\dech}{de}
\DeclareMathOperator{\child}{ch}
\DeclareMathOperator{\family}{fam}
\DeclareMathOperator{\anch}{an}

\DeclareMathOperator{\dg}{deg_{\mathcal{G}}}

\DeclareMathOperator{\x}{\boldsymbol{\lambda}_\V}
\DeclareMathOperator{\m}{\mu_\V}
\DeclareMathOperator{\dpath}{\stackrel{\rightarrow}{\pi}}
\DeclareMathOperator{\dwp}{\stackrel{\rightarrow}{\wp}}
\DeclareMathOperator{\bch}{\flat^{out}}
\DeclareMathOperator{\bpa}{\flat^{in}}

\def\V{\mathcal{V}}
\def\E{\mathcal{E}}
\def\G{\mathcal{G}}

\begin{document}

\def\spacingset#1{\renewcommand{\baselinestretch}%
{#1}\small\normalsize} \spacingset{1}


\if0\blind
{
  \title{\bf  Second-order and local characteristics of network intensity functions}
  \author{Matthias Eckardt\thanks{Financial support from the Spanish Ministry of Economy and Competitiveness via grant MTM 2013-43917-P is gratefully acknowledged.}\hspace{.2cm}\\
    Department of Computer Science, Humboldt Universit\"{a}t zu Berlin, Berlin, Germany\\
    and \\
   Jorge Mateu\\
    Department of Mathematics, University Jaume I, Castell\'{o}n, Spain}
  \maketitle
} \fi

\if1\blind
{
  \bigskip
  \bigskip
  \bigskip
  \begin{center}
    {\LARGE\bf Title}
\end{center}
  \medskip
} \fi

\bigskip
\begin{abstract}
The last decade has witnessed an increase of interest in the spatial analysis of structured point patterns over networks whose analysis is challenging because of geometrical complexities and unique methodological problems. In this context, it is essential to incorporate the network specificity into the analysis as the locations of events are restricted to areas covered by line segments. Relying on concepts originating from graph theory, we extend the notions of first-order network intensity functions to second-order and
local network intensity functions. We consider two types of local indicators of network association functions which can be understood as adaptations of the primary ideas of local analysis on the plane. We develop the node-wise and cross-hierarchical type of local functions. A real dataset on urban disturbances is also presented.
\end{abstract}

\noindent%
\vfill

\newpage
\spacingset{1.45} 

\section{Introduction}

The statistical analysis of spatial point patterns and processes is a highly attractive field of applied research across many disciplines studying the spatial arrangement of coordinates of events in planar spaces, in the sphere or over networks. Apart from point patterns in planar spaces or the sphere, the last decade witnessed an enormous increase of interest in the spatial analysis of structured point patterns and event driven data over network domains. Various extensions of classical spatial domain statistics to the network space have been proposed, which in turn rely on mathematical graph theory. For example,  \cite{GEAN:GEAN341} extended the Clark-Evans statistics to point patterns over planar network, \cite{Okabe:Yamada:2001} introduced a generalization of Ripley's $K$-function \citep{ripley:76} to the network domain, \cite{Shino2008} and \cite{SheZhuYeEtAl2015} proposed generalized Voronoi diagrams for network data, and \cite{ShiodeShiode2011} discussed the applicability of network-based and ordinary kriging techniques for street-level interpolation. Most recently, \cite{Anderes2017} covered parametric classes of covariance functions and \cite{baddeley2017} discussed second-order pseudostationary type of spatial point patterns over networks. In point patterns over networks, the positions of events are pre-configured by a set of line segments (e.g. roads) connecting pairs of fixed planar locations. In other words, treating the line segments as edges and the planar locations as nodes of an arbitrarily shaped graph, this implies that the positions of an event is governed by a geometric structure such that the point pattern can only be observed upon the edges contained in the network.

To date, a huge range of methodological and also applied papers covering global  characteristics of spatial point patterns over networks exist. Among these papers, various extensions of kernel density smoothers and second-order moment measures and functions have been proposed including the work of \cite{Borruso:2005,Borruso:2008}, \cite{McSwiggan2017}, \cite{MoradiRodriguez-CortesMateutoappear}, \cite{Ni2016}, \cite{Okabe:Satoh:2009}, \cite{Okabe:Sugihara:2012} and \cite{Yu2015}.  \cite{Ang2010},  \cite{Ang:Baddeley:Nair2012}, \cite{Baddeley:Jammalamadaka:Nair:2014} and \cite{Spooner2004} focussed on generalizations of Okabe's and Yamada's network $K$-function \citep{Okabe:Yamada:2001} controlling for the geometry of the network.  A thorough discussion of the impact of different network structures on network-based extensions of Ripley's $K$-function is given in \cite{LAmb2016}. Similar to the analysis of classical spatial point patterns, most of these contributions focused on the exploration and description of interrelations among events over the network and the underlying characteristics of the observed spatial point pattern. 

Although less frequently, several authors also considered autocorrelation between lagged edges or nodes by means of network distances. Similar to spatial autocorrelation statistics, network autocorrelation statistics express associations among measurements attributed to nodes over a network. Early adaptations of spatial autocorrelation statistics to networks have been presented by \cite{ErbringYoung1979} and \cite{DoreianTeuterWang1984} with respect to social networks, and also \cite{black1992}, who applied Moran's $I$ statistic to model autocorrelations of flow data over planar networks.  Further contributions that cover  autocorrelation functions for planar networks are the papers by \cite{Chun2008} and  \cite{Chun2013}. An in-depth treatment of network autocorrelation is given in \cite{GEAN:GEAN773}.

Lastly, several authors dealt with the analysis of local characteristics and covered   clustering and hotspot detection over planar networks. Early contributions to the local analysis of point patterns over spatial networks are \cite{Roger1996} who discussed a local cluster detection based on a $\chi^2$ test, and  \cite{Shino2004} who proposed a network cell count method. Further contributions to clustering and hotspot detection over planar networks include the $L$-function analysis \citep{LiEtAll2015}, the analysis of multiscale clusters \citep{Shino2009} as well as cluster detection over road traffic \citep{Young2014} or flow data \citep{TaoThill2016}. Adaptation of local indicators of spatial association (LISA) functions to the network domain and local $K$ functions have been discussed by \cite{Yamada2007} and  \cite{Yamada2010}. These LISA, resp. local $K$, functions have been coined local indicators of network-constrained clusters (LINCS), resp. KLINCS, by the authors. Similar approaches have been covered by \cite{Berglund1999} and \cite{FlahautMouchartMartinEtAl2003} who proposed a local $G$ statistic, and \cite{SteenberghenDufaysThomasEtAl2004} who discussed a local $I$ statistic. \cite{wang2017} applied a hierarchical Bayesian model framework for the analysis of local spatial pattern, whereas \cite{Schweitzer2006} implemented a kernel density smoother for hot spot analysis which yields to local intensity estimates.

\begin{figure}[h!]
\begin{center}
\subfloat[]{\label{Fig:disc}
\includegraphics[width=3in]{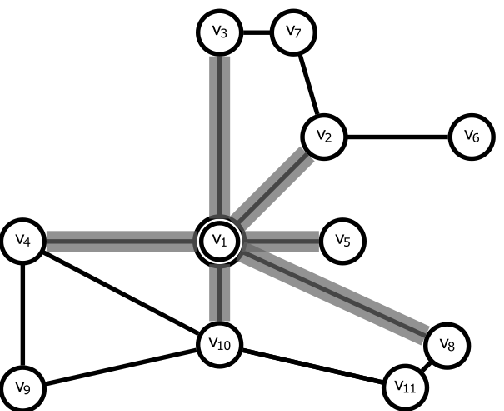}}
\subfloat[]{\label{Fig:AreaNach}
\includegraphics[width=3in]{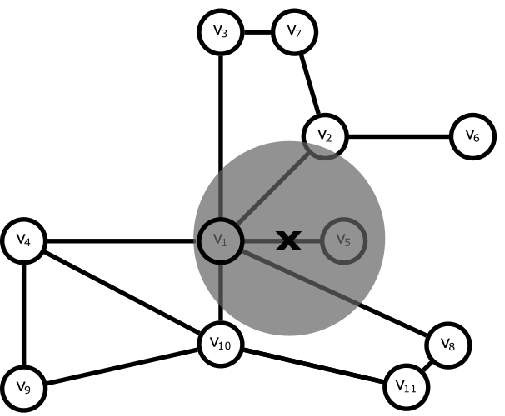}}
\caption{Examples of different areas covered by calculus in an artifical planar network: (a) edge intervals covered by the neighborhood of node $v_1$ and (b) disc of radius $r$ centred at an artifical event \textbf{X} along an edge. Vertex $v_1$ is indicated by double circles, areas covered by the edge interval covered by the neighborhood of $v_1$ and the disc centred at the origin \textbf{X} are depicted in gray.\label{Fig:areas}}
\end{center}
\end{figure}

When dealing with spatial point data collected over networks, it is essential to  incorporate the network specificity into the calculus as the locations of events are restricted to areas covered by line segments. Predominantly, as for traditional spatial point patterns statistics, techniques for analysing point patterns over spatial networks are defined with respect to pairwise metric distances between the locations of events. 
 In the most general case, this results in computations of point characteristics which only consider events within a disc of radius $r$ centred around the origin, as illustrated in Figure \ref{Fig:disc}. However, when dealing with real-world planar networks consisting of a wide variety of differently sized and differently shaped edges, circular definitions appear to be less suitable to incorporate the  characteristics and specificity of the graph. 

An alternative formalism, which in turn relies on concepts originating from graph theory,  has recently been introduced by \cite{Eckardt2016b} and is illustrated in Figure \ref{Fig:AreaNach}. Here, different from metric distance approaches, the distance boundaries used for calculation are determined exclusively by the inherent network elements independently of the length of the edges, e.g. all edge intervals contained in the neighborhood. In detail, \cite{Eckardt2016b} defined a class of network intensity functions and various intensity-based statistics for differently shaped graphs and various levels of aggregation covering undirected, directed and also partially directed networks. By this, various intensity measures have been defined for different types of networks, either with respect to events that occur on edges joining pairs of nodes, to neighboring nodes or vertex-edge sequences in form of paths and trails. 

Although this approach provides additional informations for point patterns over spatial networks, second-order or local characteristics of network intensity functions have not been presented so far. To address these limitations, we propose extensions of the network intensity formalism with respect to second-order characteristics and discuss adaptations of LISA functions to network intensity functions.  To provide a clearer classification in context, we denote these new LISA functions as local indicators of network association (LISNA). We note that the second-order analysis of point patterns over spatial network has recently also been addressed by \cite{Rakshit2017} who considered different metric distances. However, the approach presented in this paper differs from this reference in many important aspects. Essentially, it covers the second-order analysis of different entities contained in the network, namely edges, subsets of vertices such a neighborhoods, and paths and omits any statements in terms of radii. In addition, the present paper establishes a link to spatial autoregression statistics and LISA functions.       

The remainder of this paper is organized as follows. A motivation and introduction  to second-order characteristics for spatial networks is given in Section \ref{Sec:2ndOrderMeasures}, whereas a discussion of weighting matrices and local characteristics for spatial networks follows in Section \ref{sec:lisa}. Applications of local Moran $I$ and local $G$ statistics to urban disturbances-related spatial network data is given in Section \ref{sec:apl}. Finally, the concluding Section \ref{sec:conclusion} comments on the major results and impacts on future research.

\section{First- and second-order characteristics of network intensity functions}\label{Sec:2ndOrderMeasures}

Before discussing second-order characteristics of network intensity functions in detail, some notation and terminology is introduced. For an in-depth treatment of graph theory, we refer the interested reader to the monographs of \cite{Bondy2008} and also \cite{Diestel2010}.

\subsection{Notation and terminology}\label{sec:graphterms}

We consider a graph $\G$ as a pair of two finite sets, vertices $\V$ and edges $\E$. The terms network and graph are used interchangeably. The shape of $\G$ could be undirected, directed or partially directed such that pairs of vertices in $\V$ are linked by at most one edge, namely a line or an arc. In general, elements of $\V$ and $\E$ will be expressed in lower cases.  

For a given network certain sets are of interest and are intensively used in the remainder of this paper. Any pair of distinct vertices which is linked by an edge is called adjacent. In this case, the vertices are termed the endpoints of an edge and the edge is incident to its endpoints. The set of all vertices which are joined by an undirected edge to node $v_i$ is termed the neighborhood of $v_i$, $\nach\left(v_i\right)=\lbrace v_j: (v_i,v_j)\in\E\rbrace$ and the number of distinct vertices contained in $\nach(v_i)$ is the degree of $v_i$ ($\dg(v_i))$. Similarly, for any directed graph, we define the parents $\pa(v_i)$, resp. children $\child(v_i)$ of $v_i$, as the set of nodes pointing to $v_i$, resp. with root $v_i$. Taking the union over both sets results in the family which will be expressed by $\family(v_i)$.  
Analogously to $\dg(v_i)$, we express the number of distinct parents of $v_i$ by $\dg^-(v_i)$ and the number of distinct children of $v_i$ by $\dg^+(v_i)$. For partial directed graphs, the degree over directed and undirected edges is   
\[
\dg^{cg}=\sum_{i\in\nach(j)\cup\family(j)}e(v_i,v_j).
\] 
 
A path is any sequence of distinct nodes and edges, and any nodes $v_i$ and $v_j$ which are joined by a path $\pi_{ij}$ are called connected. If all edges along a path   are directed, the path is called directed path where we assume that the path is direction preserving. That is, we do not consider sequences of directed edges in which a head-to-head or tail-to-tail configuration exists. A directed path from $v_i$ to $v_j$  will be indicated by $\dpath_{ij}$.  In addition, we call any vertex $v_i$ pointing to $v_j$ an ancestor of $v_j$ and write $\anch(v_j)=\lbrace v_i\in\dpath_{ij-1}\rbrace$ to denote the set of ancestors of $v_j$. Similarly, we say that $v_j$ is a descendant of $\lbrace v_i\rbrace$ if $\anch(v_j)=\lbrace v_i\rbrace$. The set of descendants of $v_i$ is indicated by $\dech(v_i)$.

\subsection{Motivation}

\begin{figure}[h!]
\begin{center}
\subfloat[]{\label{Fig:1a}
\includegraphics[scale=.75]{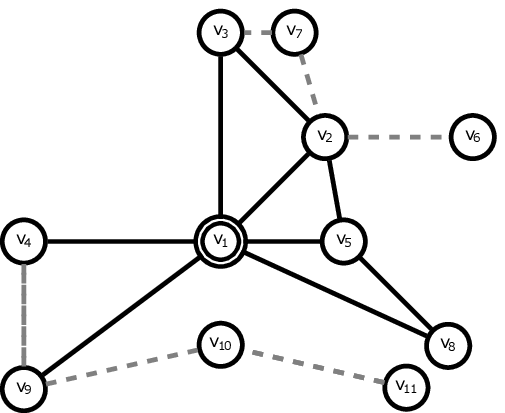}}
\subfloat[]{\label{Fig:1b}
\includegraphics[scale=.75]{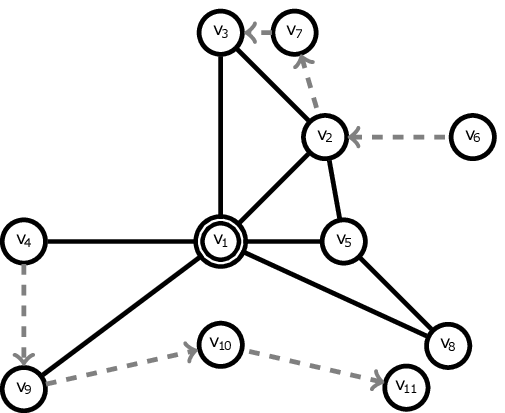}}
\subfloat[]{\label{Fig:1c}
\includegraphics[scale=.75]{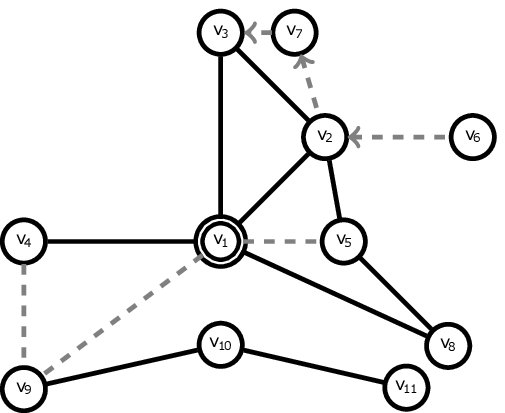}}
\caption{Examples of possible pairs of paths in an artificial network indicated by using dashed lines: (a) two undirected paths, (b) two diametrically directed paths and (c) a mixed pair of paths consisting of one undirected and one directed path.}
\label{Fig:1}
\end{center}
\end{figure}

For motivation, we consider an arbitrarily shaped spatial network with vertices $v_1$ to $v_{11}$ and a set of edges joining some, but not all pairs of vertices, as depicted in Figure \ref{Fig:1}. For simplicity, assume Figure \ref{Fig:1} displays a traffic network such that edges correspond to roads and vertices correspond to segmenting entities such as crossings. Typically, certain roads are unidirectional by nature such that traffic can only flow in one direction while other roads remain bidirected. That is, our spatial network contains directed as well as undirected edges and movements along the network appear as a sequence of either directed or undirected edges. However, alternative sequences might also be present in real-world spatial networks and corresponding sequences could easily be defined. Despite such heterogeneity,  some roads might also be affected by speed limits such that movements along such network sections is decelerated.

Given a spatial point pattern over a traffic network, one could be interested in the description of egdewise, nodewise or pathwise characteristics such as the number of events that felt onto a specific road segment or took place within a certain neighborhood structure or along a path. Such characteristics have been addressed by \cite{Eckardt2016b} by means of edgewise, node-wise and path-wise counting measures, first-order intensity functions as well as various $K$-functions for directed, undirected and mixed networks.  Despite first-order characteristics, one might also be interested in the variation or association between pairs of edges, neighborhood structures or paths. However, considering two disjoint edges, neighborhood structures or paths,  multiple second-order characteristics can be defined addressing either similar or diverse shapes. That is, the second-order edgewise intensity function could either refer to pairs of directed edges, pairs of undirected edges or,  alternatively, consider pairs of one directed and of one undirected edge. In addition, pairs of directed edges might also have a diametrical orientation in the network. Similarly, for second-order neighborhood characteristics, one  could be interested in the characterization of events that fell into two neighborhoods in case of undirected or mixed networks, or consider either two sets of parents, two sets of children or only one set of parents and one set of children. In addition, for higher-order neighborhood structures, one could also consider pairs of ancestors or descendants.

An illustration of three different pairs of paths is shown in Figure \ref{Fig:1}.  Figure \ref{Fig:1a} highlights two undirected paths joining $v_3$ to $v_6$ and $v_4$ to $v_{11}$. In contrast, two diametrically shifted paths are shown in Figure \ref{Fig:1b}. Finally, Figure \ref{Fig:1c} contains one undirected path ($v_9$ to $v_5$) and one directed path ($v_6$ to $v_3$). For any of these paired paths, one might be interested in the expected number of points, the variation in number of points or the correlation between the number of points that felt onto both paths. In addition, as for classical spatial point patterns, one could also be interested in the probability of an event in path $a$ given an event in path $b$. 

At the same time, one might also be interested in the interrelation between pairs of edgewise and path-wise intensity functions or in the contribution of a specific node-wise intensity function computed at a distinct node to the distribution of all node-wise intensity functions over the network.

\subsection{Recapitulating first-order network intensity functions}\label{sec:firstoder}

\begin{figure}[h!]
\begin{center}
\subfloat[]{\label{Fig:edge}
\includegraphics[scale=1.05]{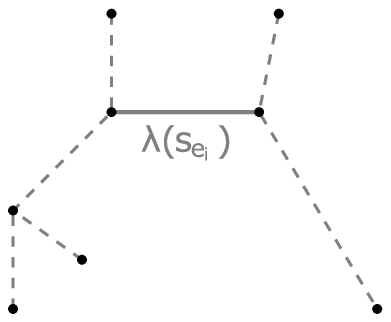}}
\subfloat[]{\label{Fig:nach}
\includegraphics[scale=1.05]{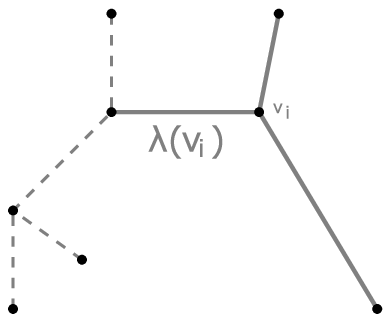}}
\subfloat[]{\label{Fig:path}
\includegraphics[scale=1.05]{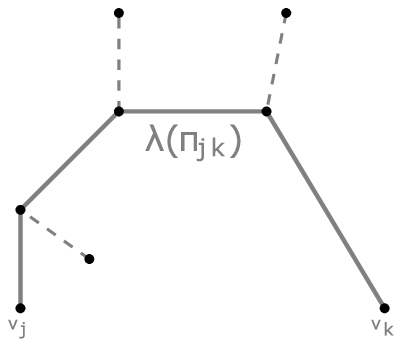}}
\caption{Illustration of different network intensity functions in undirected networks: edge intervals used for calculation of (a) the edgewise, (b) the node-wise and (c) the path-wise intensity function. Corresponding edge intervals used for computation are highlighted in solid gray lines.}
\label{Fig:UGnetintensity}
\end{center}
\end{figure}

Before we discuss the second-order statistics for spatial networks, we briefly present the basic ideas of counting measures and statistics with respect to points contained in $\mathcal{S}_{E(G)}$ for different types of networks and recapitulate different first-order network intensity functions. Here, we first treat undirected graphs and discuss directed and partially directed graphs consecutively. Extension to higher-order characteristics are straightforward and follow naturally as generalizations of well-known point pattern characteristics. In general, three different types of network intensity functions can be addressed referring to different levels of network resolutions. These are the edgewise, the node-wise and the path-wise intensity functions. All three types of network intensity functions are illustrated in Figure \ref{Fig:UGnetintensity} with respect to undirected networks.

Following the ideas and notation of \cite{Eckardt2016b}, we  address the set of nodes at fixed locations $\mathbf{s}_{v}=(\mathbf{x}_{v},\mathbf{y}_{v})$ contained in a spatial network $\G$ by $\V_s(\G)$ and refer to the set of edge intervals connecting pairs of fixed locations in $\G$ by $\mathcal{S}_{E_s(\G)}=\lbrace s_{e_1},\ldots,s_{e_k}\rbrace$.  In addition, we express the locations of a point process $X(\tilde{s})$ over $\mathcal{S}_{E_s(G)}$ by $\tilde{\mathbf{s}}=(\tilde{\mathbf{x}}, \tilde{\mathbf{y}})$. The location of node $v_i$ is $s_{v_i}=(x_{v_i},y_{v_i})$. Clearly, under this definition, point patterns are only allowed to occur within a given edge interval contained in $\G$. That is, the locations $\tilde{\mathbf{s}}$ are said to occur randomly within edge intervals spanned between any two fixed locations $s_{v_i}$ and $s_{v_j}$ of $\mathbf{s}_{v}$, e.g. on road segments.  By this, we understood a path as a sequence of consecutive edge intervals and the distance $d_\G(v_i,v_j)$ between any two nodes in $\V_s(\G)$ is the number of consecutive edges joining $v_i$ and $v_j$, that is the length of a path. Hence, the shortest path distance is the minimum number of consecutive edges needed to move from $v_i$ to $v_j$ along a network.

We explicitly note, that these definitions lead to fundametally different concepts of  lenght as considered in \cite{Ang:Baddeley:Nair2012}, \cite{Baddeley:Jammalamadaka:Nair:2014},  and \cite{baddeley2017} who defined the length of a path as the sum over Euclidean distances between consecutive nodes contained in a path, and in \cite{Rakshit2017} who also considered alternative metric distances. By this, the shortest path distance is the minimum of metric distance totals of all paths joining two locations and it is not defined as the minimum number of traversed edges along a path.

\subsubsection{First-order network intensity functions for undirected networks}\label{subsec:undirected}

Let $N(s_{e_i})$ be the number of points that fall into the undirected edge interval $s_{e_i}$ and $ds_{e_i}$ denote an infinitesimal intervals containing $s_{e_i}$ such that $N(ds_{e_i})=N(s_{e_i}+ds_{e_i})-N(s_{e_i})$. Then, we have for the first-order edgewise intensity function 
\begin{equation}
\label{eq:intensityUG}
\lambda(s_{e_i})=\lim_{|ds_{e_i}|\rightarrow 0}\left\{\frac{\mathds{E}\left[N(ds_{e_i})\right]}{|ds_{e_i}|}\right\}, s_{e_i}\in\mathcal{S}_{E(G)}.
\end{equation}
Using this expression, we obtain the node-wise mean intensity function $\lambda(v_i)$ for any given node  $v_i$ contained in $\G$ by averaging \eqref{eq:intensityUG} over the set of adjacent nodes. That is,  
\[
\lambda(v_i)=\frac{1}{|\dg(v_i)|}\sum_{v_j\in\nach(v_i)}\lambda(s_{e_i}). 
\]

Besides, apart from any such average intensities of points per neighborhood, one can  define neighborhood intensity functions using the set of incident edges. To this end, let $\flat(v_i)$ denote the set of edge intervals with endpoint $v_i$, $N(\flat(v_i))$ be the number of points in $\flat(v_i)$ and $d\flat(v_i)$ denote an infinitesimal area covering $\flat(v_i)$. By this, we define the non-averaged neighborhood intensity function $\lambda(\nach(v_i))$ as
\begin{equation}
\label{eq:intensityNE}
\lambda(\nach(v_i))=\lim_{|d\flat(v_i)|\rightarrow 0}\left\{\frac{\mathds{E}\left[N(d\flat(v_i))\right]}{|d\flat(v_i)|}\right\}.
\end{equation}

Using the same ideas as for $\lambda(v_i)$ and $\lambda(\nach(v_i))$, we can define an averaged and a non-averaged version of pathwise intensity functions for a path $\pi_{ij}$ joining $v_i$ to $v_j$. The average pathwise intensity function $\lambda(\pi_{ij})$, which has been introduced by \cite{Eckardt2016b}, is given as 
\[
\lambda(\pi_{ij})=\frac{1}{|\mathcal{N}_\pi|}\sum_{v_j\in\pi_{ij}}\lambda(s_{e_i})
\]
where $\mathcal{N}_\pi$ is the cardinality of consecutive edge intervals traversed in $\pi_{ij}$. As for \eqref{eq:intensityNE}, writing $\wp_{ij}$ to denote the set of edge intervals traversed once along $\pi_{ij}$, we define the non-average path-wise intensity function as
\begin{equation}\label{eq:pathwiseNonAveUG}
\lambda(\pi^*_{ij})=\lim_{|d\wp_{ij}|\rightarrow 0}\left\{\frac{\mathds{E}\left[N(d\wp_{ij})\right]}{|d\wp_{ij}|}\right\}
\end{equation}
where $N(d\wp_{ij})=N(\wp_{ij}+d\wp_{ij})-N(\wp_{ij})$ and $|d\wp_{ij}|$ is the area contained in $d\wp_{ij}$.     

\subsubsection{First-order network intensity functions for directed networks}\label{subsec:directed}

\begin{figure}[h!]
\begin{center}
\subfloat[]{\label{Fig:in}
\includegraphics[scale=1.05]{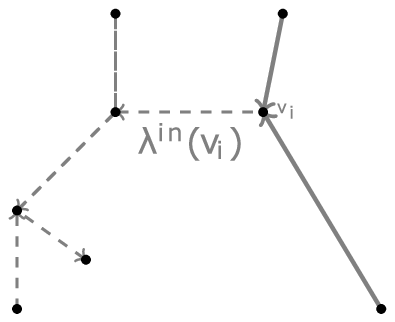}}
\subfloat[]{\label{Fig:out}
\includegraphics[scale=1.05]{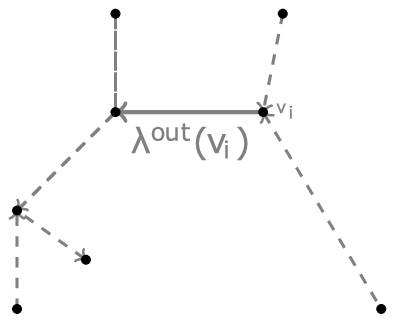}}
\subfloat[]{\label{Fig:dipath}
\includegraphics[scale=1.05]{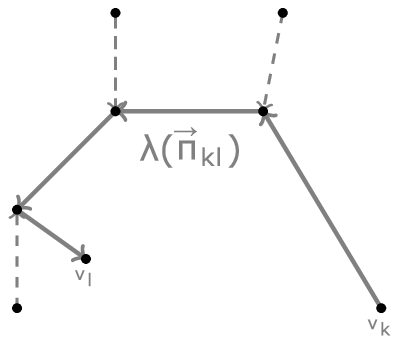}}
\caption{Illustration of different network intensity functions in directed networks:  edge intervals considered by (a) parent-wise mean, (b) children-wise mean and (c) directed pathwise intensity functions. Corresponding edge intervals used for computation are highlighted in solid gray lines.}
\label{Fig:DGnetintensity}
\end{center}
\end{figure}

To cover directed graphs, slightly modifications of the previous notations are required. To this end, let $N(s_{e_i}^{in})$ express the number of events on an edge leading to and $N(s_{e_i}^{out})$ be the number of events on an edge departing from a vertex of interest,  and $ds_{e_i}^{in}$ and $ds_{e_i}^{out}$ denote infinitesimal intervals containing $s_{e_i}^{in}$ and $s_{e_i}^{out}$. By using this formal set-up, different levels of network resolutions can be addressed, each containing two different types of directed network intensity functions due to the direction of the edges. Figure \ref{Fig:DGnetintensity} illustrates the edge intervals used for computation of parent- and children-wise intensity functions and also the directed path-wise intensity function.  

Substitution of $N(s_{e_i}^{in})$ or $N(s_{e_i}^{out})$ for $N(s_{e_i})$ in \eqref{eq:intensityUG} yields to the directed first-order edgewise intensity functions, 
\begin{equation}
\label{eq:intensityparent}
\lambda(s_{e_i}^{in})=\lim_{|ds_{e_i}^{in}|\rightarrow 0}\left\{\frac{\mathds{E}\left[N(ds_{e_i}^{in})\right]}{|ds_{e_i}^{in}|}\right\}, s_{e_i}^{in}\in\mathcal{S}_{E(G)}
\end{equation}
and 
\begin{equation}
\label{eq:intensitychild}
\lambda(s_{e_i}^{out})=\lim_{|ds_{e_i}^{out}|\rightarrow 0}\left\{\frac{\mathds{E}\left[N(ds_{e_i}^{out})\right]}{|ds_{e_i}^{out}|}\right\}, s_{e_i}^{out}\in\mathcal{S}_{E(G)}.
\end{equation}
As for the undirected case, we obtain the parent-wise mean intensity function by averaging of \eqref{eq:intensityparent} over the set of parents of $v_i$, 
\[
\lambda^{in}(v_i)=\frac{1}{|\dg^+(v_i)|}\sum_{v_j\in\pa(v_i)}\lambda(s^{in}_{e_i}) 
\]
and the children-wise mean intensity function by averaging of \eqref{eq:intensitychild} over the set of children of $v_i$,
\[
\lambda^{out}(v_i)=\frac{1}{|\dg^-(v_i)|}\sum_{v_j\in\child(v_i)}\lambda(s^{out}_{e_i}).
\]

As in the undirected case, one can define non-averaging version of $\lambda^{in}(v_i)$ and $\lambda^{out}(v_i)$ with respect to the sets of incident edge intervals with head or tail $v_i$, namely incident  edge intervals pointing to $v_i$ ($\lambda(\pa(v_i))$) and incident  edge intervals departing from $v_i$ ($\lambda(\child(v_i))$). Defining $\bpa(v_i)$ (resp. $\bch(v_i)$) as the set of edge intervals pointing to (resp. departing from) $v_i$ and using the same terminology as before, we obtain the non-averaging parent-wise (resp. children-wise) intensity function by substituting $N(d\bpa(v_i))$ (resp. $N(d\bch(v_i))$) for $N(d\flat(v_i))$ and $d\bpa(v_i)$ (resp. $d\bch(v_i)$) for $d\flat(v_i)$ in \eqref{eq:intensityNE}.

Extensions of path-wise intensity functions to directed networks follow naturally as a generalization of $\lambda(\pi_{ij})$. For the directed path $\dpath_{ij}$ pointing to $v_j$
we have 
\begin{equation}\label{eq:diMeanpath}
\lambda(\dpath_{ij})=\frac{1}{|\mathcal{N}_{\dpath}|}\sum_{v_i\in\dpath_{ij}}\lambda(s_{e_i})
\end{equation}
where $\mathcal{N}_{\dpath}$ is the cardinality of consecutive edge intervals along $\dpath_{ij}$. We note that in general, different from nodewise calculations,  \eqref{eq:diMeanpath} is defined for an ordered pair of endpoints of a directed path such that $\lambda(\dpath_{ij})$ and $\lambda(\dpath_{ji})$ refer to different sequences of edge intervals contained in $\G$. By this, as $\lambda(\dpath_{ij})\neq\lambda(\dpath_{ji})$, it suffices to consider only one expression to refer to either paths with root or with head $v_i$. Besides,  two distinct paths $\dpath_{ij}$ and $\dpath_{jk}$ are allowed to have a common endpoint, such that $v_j$ serves as a terminus in $\dpath_{ij}$ and as a root of $\dpath_{jk}$.
Apart from \eqref{eq:diMeanpath}, we define the directed non-average path-wise intensity function $\lambda(\dpath^*_{ij})$ as
\begin{equation}\label{eq:dipathnonaverage}
\lambda(\dpath^*_{ij})=\lim_{|d\dwp_{ij}|\rightarrow 0}\left\{\frac{\mathds{E}\left[N(d\dwp_{ij})\right]}{|d\dwp_{ij}|}\right\}
\end{equation}
where $\dwp_{ij}$ is the set of edges intervals traversed once along a directed path with root $v_i$ and head $v_j$, $d\dwp_{ij}$ is an infinitesimal interval contained in $\dwp_{ij}$ and $|d\dwp_{ij}|$ is the area covered by $d\dwp_{ij}$.

Apart from directed path-wise intensity functions, one could also consider all edge intervals along a directed path except its origin or terminus. That is, the information contained in the ancestors or descendants of a distinct node. Writing $\wp^{-i}_{ij}$ for the set of edge intervals contained in $\dech(v_i)$ and $\wp^{-j}_{ij}$ for the set of edge intervals contained in $\anch(v_j)$, a modification of \eqref{eq:dipathnonaverage} yields to
\[
\lambda(\anch(v_j))=\lim_{|d\wp^{-j}_{ij}|\rightarrow 0}\left\{\frac{\mathds{E}\left[N(d\wp^{-j}_{ij})\right]}{|d\wp^{-j}_{ij}|}\right\}
\]
and
\[
\lambda(\dech(v_j))=\lim_{|d\wp^{-j}_{ji}|\rightarrow 0}\left\{\frac{\mathds{E}\left[N(d\wp^{-j}_{ji})\right]}{|d\wp^{-j}_{ji}|}\right\}.
\]

\subsubsection{First-order network intensity functions for partially directed networks}

As partially directed networks are defined as hybrids of directed and undirected networks, we obtain various types of network intensity function as union over the directed and the undirected intensity functions. Using the results of Sections \ref{subsec:undirected} and \ref{subsec:directed}, we obtain the node-wise mean intensity function for partial networks $\lambda^{cg}(v_i)$ by    
\begin{equation}
\label{eq:intensitychain}
\lambda^{cg}(v_i)=\frac{1}{|\dg^{cg}(v_i))|} \lambda^{out}(v_i)\cup \lambda^{in}(v_i)\cup \lambda(v_i).
\end{equation}
Alternative versions of \eqref{eq:intensitychain} follow naturally by modification of the union sets. For example, the union $\pa(\cdot)\cup\child(\cdot)$ would only consider directed adjacent edges, whereas the union $\nach(\cdot)\cup\child(\cdot)$ will exclude any edge pointing to a node of interest. Using the previous results for directed and undirected networks,  we can define non-average versions of \eqref{eq:intensitychain} as unions over $\lambda(\nach(v_i)),\lambda(\pa(v_i))$ and $\lambda(\child(v_i))$ such as the family-wise intensity function 
\[
\lambda^{cg}(\family(v_i))=\lambda(\pa(v_i))\cup \lambda(\child(v_i))
\]
which expresses the expected number of counts along all directed edge intervals which are incident to node $v_i$.

\subsection{Second-order intensity and covariance density functions for planar networks}\label{sec:2ndorder}

Having point patterns over spatial networks under study, one could be interested in the variation of intensity functions among two different graph entities, e.g. the pairs of distinct edges, neighborhoods or paths contained in the graph. For classical point pattern statistics, such variations are usually expressed by means of second-order properties of the point pattern such as the second-order intensity or the auto- and cross-covariance density functions. This section covers extensions of both functions to pairs of distinct edge intervals, pairs of distinct node-wise sets of edge intervals or pairs of sequences of edge intervals contained in spatial networks. These functions can then be used to characterize the locations of events over the spatial network, which in turn 
could exhibit randomness, clustering or regularity.

\subsubsection{Edgewise second-order intensity and covariance density functions}

Consider $s_{e_i}$ and $s_{e_j}$ denote two distinct edge intervals of possibly different shape or length contained in $\G$. Then, for any distinct edge intervals contained in any such pair, we can define either directed or undirected counting measures. First, assume that $\G$ is undirected. Then, using the same notation as before, we obtain the second-order edgewise intensity function $\lambda(s_{e_i}, s_{e_j})$ as   
\begin{equation}
\label{eq:SecondOrderIntensity}
\lambda(s_{e_i}, s_{e_j})=\lim_{|ds_{e_i}, ds_{e_j}|\rightarrow 0}\left\{\frac{\mathds{E}\left[N(ds_{e_i}), N(ds_{e_j})\right]}{|ds_{e_i}\times ds_{e_j}|}\right\}
\end{equation}
where $s_{e_i}\neq s_{e_j}$. Less formally, $\lambda(s_{e_i}, s_{e_j})$ is the expected number of counts for pairs of distinct undirected edge intervals. However, although \eqref{eq:SecondOrderIntensity}
 can be used to define edgewise versions of Ripleys' $K$-function \citep{ripley:76}, it does not provide a suitable characterization of the  theoretical properties of the spatial point pattern. An alternative second-order characteristic which better describes the theoretical properties of the spatial point pattern is the edgewise covariance density function $\gamma(s_{e_i}, s_{e_j})$, 
\[
\gamma(s_{e_i}, s_{e_j})=\lambda(s_{e_i}, s_{e_j})-\lambda(s_{e_i})\lambda(s_{e_j}).
\]
  
As discussed in Section \ref{sec:firstoder}, several different second-order edgewise intensity and covariance functions can be defined. An overview of second-order edgewise intensity functions and edgewise auto-covariance functions which can be defined for directed, undirected and partially directed networks is given in Table \ref{Tab:edgeintens}.

\begin{table}
\caption{Edgewise second-order intensity functions\label{Tab:edgeintens}} 
\begin{center}
  \begin{tabular}{rrrr}
 type of network & $2^{nd}$ order intensity &  auto-covariance & Counting measures. \\
 \hline   
undirected & $\lambda(s_{e_i}, s_{e_j})$ & $\gamma(s_{e_i}, s_{e_j})$  & $N(ds_{e_i}), N(ds_{e_j})$  \\ 
directed & $\lambda(s_{e_i}^{out}, s_{e_j}^{out})$ & $\gamma(s_{e_i}^{out}, s_{e_j}^{out})$ & $N(ds_{e_i}^{out}), N(ds_{e_j}^{out})$\\
directed & $\lambda(s_{e_i}^{in}, s_{e_j}^{in})$& $\gamma(s_{e_i}^{in}, s_{e_j}^{in})$ & $N(ds_{e_i}^{in}), N(ds_{e_j}^{in})$ \\
directed & $\lambda(s_{e_i}^{in}, s_{e_j}^{out})$& $\gamma(s_{e_i}^{in}, s_{e_j}^{out})$ & $N(ds_{e_i}^{in}), N(ds_{e_j}^{out})$\\
partially directed & $\lambda(s_{e_i}, s_{e_j}^{out})$ & $\gamma(s_{e_i}, s_{e_j}^{out})$ & $N(ds_{e_i}), N(ds_{e_j}^{out})$\\
partially directed & $\lambda(s_{e_i}, s_{e_j}^{in})$ & $\gamma(s_{e_i}, s_{e_j}^{in})$ & $N(ds_{e_i}), N(ds_{e_j}^{in})$\\
  \end{tabular}
\end{center}
\end{table}

\subsubsection{Node-wise second-order intensity and covariance density functions}

Similarly to the edgewise second-order intensity functions, we could also be interested in the characterization of variations among distinct subsets of edge intervals contained in a spatial network. For this, one could address either the pairwise variation with respect to distinct nodes such as the second-order or covariance density functions for pairs of neighbors, or the pairwise variation with respect to an identical vertex, e.g. the variation of intensities between the parents and children of a specific node. 

Given two sets of distinct neighborhoods $\nach(v_i)$ and $\nach(v_j)$ where $v_i\neq v_j$, a generalization of \eqref{eq:SecondOrderIntensity} results in
\[
\lambda(\nach(v_i),\nach(v_j))=\lim_{|d\flat(v_i),d\flat(v_j)|\rightarrow 0}\left\{\frac{\mathds{E}\left[N(d\flat(v_i)),N(d\flat(v_j))\right]}{|d\flat(v_i)\times d\flat(v_j)|}\right\}.
\]

Using the same arguments as for the edgewise second-order intensity function, we define the auto-covariance density functions as
\[
\gamma(\nach(v_i),\nach(v_j))=\lambda(\nach(v_i),\nach(v_j))-\lambda(\nach(v_i))\lambda(\nach(v_j))
\]
where $\lambda(\nach(v_i)$  and $\lambda(\nach(v_j)$ are the non-averaged node-wise intensity functions of $v_i$ and $v_j$ as defined in \eqref{eq:intensityNE}. An overview of different node-wise second-order intensity and auto-covariance functions is given in Table \ref{Tab:nodeintens}.  

\begin{table}
\caption{Node-wise second-order intensity functions where $\flat^{fam}(v_j)=\flat^{in}(v_j))\cup N(d\flat^{out}(v_j))$. \label{Tab:nodeintens}} 
\begin{center}
  \begin{tabular}{rrrr}
 type of network & $2^{nd}$ order intensity &  auto-covariance & Counting measures \\
 \hline   
undirected & $\lambda(\nach(v_i),\nach(v_j))$ & $\gamma(\nach(v_i),\nach(v_j))$  & $N(d\flat(v_i)),N(d\flat(v_j)$  \\ 
directed & $\lambda(\pa(v_i),\pa(v_j))$ & $\gamma(\pa(v_i),\pa(v_j))$ & $N(d\flat^{in}(v_i)),N(d\flat^{in}(v_j)$\\
directed & $\lambda(\child(v_i),\child(v_j))$& $\gamma(\child(v_i),\child(v_j))$ & $N(d\flat^{out}(v_i)),N(d\flat^{out}(v_j)$ \\
directed & $\lambda(\pa(v_i),\child(v_j))$& $\gamma(\pa(v_i),\child(v_j))$ & $N(d\flat^{in}(v_i)),N(d\flat^{out}(v_j)$ \\
directed & $\lambda(\family(v_i),\family(v_j))$& $\gamma(\family(v_i),\family(v_j))$ & $N(d\flat^{fam}(v_i)),N(d\flat^{fam}(v_j)$ \\
partially directed & $\lambda(\nach(v_i), \pa(v_i))$ & $\gamma(\nach(v_i), \pa(v_i)$ & $N(d\flat(v_i)),N(d\flat^{in}(v_i)$\\
partially directed & $\lambda(\nach(v_i), \child(v_i))$ & $\gamma(\nach(v_i), \child(v_i))$ & $N(d\flat(v_i)),N(d\flat^{out}(v_i)$\\
partially directed & $\lambda(\nach(v_i),\family(v_j))$& $\gamma(\nach(v_i),\family(v_j))$ & $N(d\flat(v_i)),N(d\flat^{fam}(v_j)$ \\
  \end{tabular}
\end{center}
\end{table}

\subsubsection{Path-wise second-order intensity and covariance density functions}

Lastly, we can also consider the variations among distinct pairs of paths contained in a network. In general, any such variation can be defined for pairs of paths with either common or different endpoints such as $V$-structures in form of $\pi_{ij}$ and $\pi_{ik}$, inverse $V$-structures in form of $\pi_{ij}$ and $\pi_{hj}$, elliptic $O$-structures in form of $\pi^{(1)}_{ij}$ and $\pi^{(2)}_{ij}$ where any edge interval is only allowed to traversed once in either $\pi^{(1)}_{ij}$ or in $\pi^{(2)}_{ij}$, or in form of two distinct paths $\pi_{ij}$ and $\pi_{kl}$. 

In general, for $\pi^*_{ij}$ and $\pi^*_{kl}$ and adopting the same ideas as before, we have
\[
\lambda(\pi^*_{ij},\pi^*_{kl})=\lim_{|d\wp_{ij},d\wp_{kl}|\rightarrow 0}\left\{\frac{\mathds{E}\left[N(d\wp_{ij}),N(d\wp_{kl})\right]}{|d\wp_{ij}\times d\wp_{kl}|}\right\}
\]
and 
\[
\gamma(\pi^*_{ij},\pi^*_{kl})=\lambda(\pi^*_{ij},\pi^*_{kl})-\lambda(\pi^*_{ij})\lambda(\pi^*_{kl})
\]
where $\lambda(\pi^*_{ij})$ and $\lambda(\pi^*_{kl})$ are non-averaged path-wise first-order intensity functions as introduced in \eqref{eq:pathwiseNonAveUG}. 

As for the edgewise and node-wise second-order characteristics, various types of path-wise second-order intensity and auto-covariance functions can easily been introduced, see Table \ref{Tab:pathintens} for a detailed list. We remark that differently from edgewise or node-wise calculations, the second-order path-wise properties either include or exclude the endpoint of a directed path such that $\lambda(\pi^*_{ij},\pi^*_{kj})\neq \lambda(\dech_{ji},\anch_{ij})$. That is, while the edge interval $s_{e_j}=(v_{j-1},v_j)$ is included by $\dpath^*_{ij}$, it is excluded by $\anch(v_j)$ as $\anch(v_j)$ only considers all edge interval along the path $\dpath^*_{ij-1}$. 

\begin{table}
\caption{Path-wise second-order intensity functions.\label{Tab:pathintens}} 
\begin{center}
  \begin{tabular}{rrrr}
 type of network & $2^{nd}$ order intensity &  auto-covariance & Counting measures \\
 \hline   
undirected & $\lambda(\pi^*_{ij},\pi^*_{kl})$ & $\gamma(\pi^*_{ij},\pi^*_{kl})$  & $N(d\wp_{ij}),N(d\wp_{kl})$  \\ 
directed & $\lambda(\dpath^*_{ij},\dpath^*_{kl})$ & $\gamma(\dpath^*_{ij},\dpath^*_{kl})$ & $N(d\dwp_{ij}),N(d\dwp_{kl})$\\

directed & $\lambda(\anch(v_j),\anch(v_k))$& $\gamma(\anch(v_j),\anch(v_k))$ & $N(d\wp^{-j}_{ij}),N(d\wp^{-k}_{lk})$ \\ 
directed & $\lambda(\dech(v_j),\dech(v_k))$& $\gamma(\dech(v_j),\dech(v_k))$ & $N(d\wp^{-j}_{ji}),N(d\wp^{-k}_{kl})$ \\ 
directed & $\lambda(\anch(v_j),\dech(v_k))$& $\gamma(\anch(v_j),\dech(v_k))$ & $N(d\wp^{-j}_{ij}),N(d\wp^{-k}_{kl})$ \\

partially directed & $\lambda(\pi^*_{ij}, \dpath^*_{kl})$ & $\gamma(\pi^*_{ij}, \dpath^*_{kl})$ & $N(d\wp_{ij}),N(d\dwp_{kl}) $\\
partially directed & $\lambda(\pi^*_{ij}, \anch(v_k))$ & $\gamma(\pi^*_{ij}, \anch(v_k))$ & $N(d\wp_{ij}),N(d\wp^{-k}_{lk})$\\
partially directed & $\lambda(\pi^*_{ij},\dech(v_k))$& $\gamma(\pi^*_{ij},\dech(v_k))$ & $N(d\wp_{ij}),N(d\wp^{-k}_{kl})$ \\
  \end{tabular}
\end{center}
\end{table}

\section{Local indicators of spatial association for network intensity functions}\label{sec:lisa}

This section introduces local indicators of network associations (LISNA) functions for network intensity which can be understood as adaptations of the primary ideas of local analysis to the analysis of spatial point patterns over a network. In general, we concern two different types of LISNA functions: node-wise LISNA functions (type 1) and  cross-hierarchical LISNA functions (type 2). While LISNA functions of type 1 are generalizations of Anselins' LISA functions \citep{Anselin1995} to node-wise intensity functions which can be applied to any global measure of spatial association, cross-hierarchical LISNA functions express the variation between individual edge intervals and different subsets of edge intervals contained in  different network entities. LISNA characteristics consider the individual contributions of a global estimator as a measure of clustering. Before we discuss LISNA functions of type 1 and of type 2 in detail, we briefly review the concept of LISA functions and related clustering approaches for the spatial domain. 

\subsection{A primer on local indicator of spatial association statistics}

Spatial cluster detection has stimulated an immense interest in efficient statistical analysis tools and several authors have contributed to this field. A local version of the Ripley's $K$-function \citep{ripley:76} has been proposed by \cite{getis:franklin:87,getis:franklin:10} in order to quantify clustering at different spatial scales. Another local statistic, the local $G$ statistic, was presented by \cite{getis:ord:92,getis:ord:10} which allows to assess the degree of spatial association at various levels of spatial refinement in an entire sample or in relation to  a single observation. \cite{stoyan:stoyan:94} introduced both local $L$- and local $g$-functions for the analysis of  neighbourhood relationships. A local $H_{i}$ statistic was introduced by  \cite{ord:getis:12} in order to measure the spatial variability while avoiding the pitfalls of using the non-spatial $F$ test for spatial data. 

We note that several authors have considered product density LISA functions for cluster detection in spatial point patterns which will not be covered here. These LISA functions originate in the papers of \cite{cressie:colins:01a,cressie:colins:01b} who considered bundles of product density LISA functions for the recognition of similarity groupings in spatial subpatterns by examing individual points in the point pattern in terms of how they relate to their adjacent points in space. Similarly, \cite{mateu:lorenzo:porcu:10} used product density LISA functions for cluster detection in the presence of substantial clutter, and \cite{moraga:montes:11} discussed the use of product density LISA functions with respect to disease clusters. 

\subsection{Local indicators of spatial network associations of type 1}

\subsubsection{Weighting matrices for spatial network intensity functions}

To discuss spatial auto-correlation and local associations among spatial point patterns over network structures, a suitable weighting matrix $\mathbf{W}$ is essential and has to be defined prior to analysis. In general, for node-wise associations, a reasonable choice of $\mathbf{W}$ is the adjacency matrix $\mathbf{A}$ of the network $\G$ which represents the structure of the graph in a compact way. However, as we will discuss next, $\mathbf{W}$ is equivalently encoded by $\mathbf{A}$ in most but not all cases. 

Most commonly, depending on the type of the network, $\mathbf{A}$ is either a symmetric or an asymmetric binary matrix of dimension $\V(\G)\times \V(\G)$. However, as we consider local associations among different sets of nodes, namely the sets of neighbors, parents or children, one might not only be interested in the association within and between different sets of nodes but also in the variation of associations for different orders of network linkages. That is, apart the graph-theoretic (first-order) definitions of neighbors, parents or children as introduced in Section \ref{sec:graphterms}, one might also be interested in the cumulative or the partial $k$-th order subset of nodes where $k=2,3,\ldots$. For notational simplicity, we will address the partial $k$-th order of $\mathbf{A}$ by adding a superscript $k$ to $\mathbf{A}$ such that $\mathbf{A}^{(k)}$ is the partial $k$-th order adjacency matrix of $\G$.  Similarly, $\mathbf{W}^{(k)}$ denotes the spatial weighting matrix of order $k$.

For $\mathbf{A}^{(1)}$, the $ij$-th element of $\mathbf{A}^{(1)}$ is only non-zero if $v_i$ and $v_j$ are joined by an edge in $\G$. Thus, if $\G$ is an undirected network, $a_{ij}=1$ also implies that $a_{ji}=1$ due to the symmetry of $\mathbf{A}$. Despite such first-order subsets of nodes,  the definition of higher-order neighborhoods, parents or children requires a careful distinction  between the notions of partial and cumulative subsets of nodes. Extensions to partial  higher-order adjacency matrices $\mathbf{A}^{(k)}$ follow naturally as generalizations of $\mathbf{A}^{(1)}$ such that $a^{(k)}_{ij}$ of $\mathbf{A}^{(k)}$ is non-zero if $v_i$ and $v_j$ are joined by $k-1$ interior nodes. That is, if $v_i$ is connected to $v_j$ by a path of length $k$ and vice versa. Different from such partial sets of order $k$, cumulative sets of order $k$ consist of all neighbors, parents or children of a distinct node up to order $k$. That is, a second-order cumulative neighborhood of order $k$ of node $v_i$ can be understood as the union over the first-order and all partial second-order neighbors of $v_i$ up to order $k$. Obviously, these are all vertices along all paths of length $k$ contained in the network with origin $v_i$. Consequently, while $\mathbf{W}^{(k)}=\mathbf{A}^{(k)}$, this equivalence in general does not hold for cumulative subsets of order $k$. 

\begin{figure}[h!]
\begin{center}
\includegraphics[width=3in]{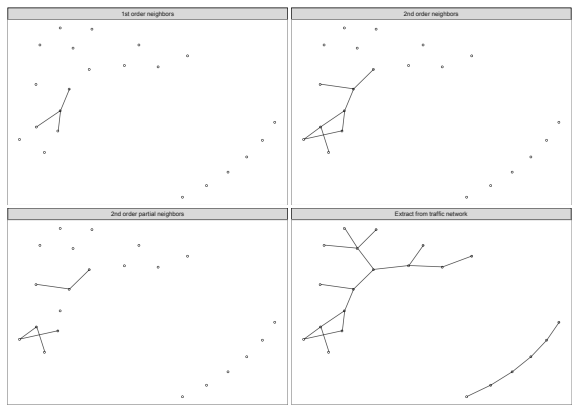}
\caption{Examples of neighboring structures in an undirected network\label{Fig:neighbors}}
\end{center}
\end{figure}

Illustrations of the first-order as well as the second-order partial and the second-order cumulative neighborhood for an undirected network are shown in Figure \ref{Fig:neighbors}. In a similar manner, we can define partial and cumulative subsets of vertices for individual nodes for directed and mixed networks. However, different from the undirected case, these sets consider the direction as well as the shape of the edges.   

\subsubsection{Node-wise LISNA functions}

We now turn to the discussion of LISNA functions of type 1. For this, let $\x$  denote a vector of dimension $n\times 1$ of either averaged or non-averaged node-wise intensity functions defined with respect to either the set of $n$ neighbors, $n$ parents, $n$ children or $n$ families contained in a network, $\m$ be the mean of $\x$ over $\G$ and $\gamma_\V^{(0)}$ the auto-covariance of $\x$. In general, we assume the type of network elements associated to $\x$ to be unique such that e.g. all elements of $\x$ are parent-wise intensity functions.  

Given the $k$-order weighting matrix $\mathbf{W}^{(k)}$ with $w^{(k)}_{ij}\neq 0$ if $v_i$ and $v_j$ are joined by a path of length $k-1$, we obtain the auto-covariance of $\x$ of order $k$ as  
\[
\gamma_\V^{(k)} = \frac{\mathbf{\x}\T \mathbf{W}^{(k)}\mathbf{\x}}{\sum_{i=1}^{n}\sum_{j=1}^{n}w^{(k)}_{ij}}.
\]

Similarly, we define the  auto-correlation of $\x$ as $\rho=\gamma^{(k)}/\gamma_\V^{(0)}$ which expresses the correlation along the network in terms of distance for different lags $k$. As for classical spatial statistics, a general approach to characterize node-wise auto-correlations is to compute auto-correlation statistics, namely Morans' $I$ statistic, Geary's $C$ statistic or Getis' and Orb's $G$ statistic.  
Assuming that $\mathbf{W}=\mathbf{A}^{(1)}$ such that $w_{ij}=1$ if and only if $v_i$ and $v_j$ are joined by an edge interval in $\G$, the Moran $I$ statistics for spatial networks is given by   
\begin{equation}\label{eq:MoranI}
I = \frac{n}{\sum_{i=1}^{n}\sum_{j=1}^{n}w_{ij}}
\frac{\sum_{i=1}^{n}\sum_{j=1}^{n}w_{ij}(\lambda(v_i)-\m)(\lambda(v_j)-\m)}{\sum_{i=1}^{n}(\lambda(v_i) - \m)^2}.
\end{equation}
and can be understood as the ratio between the product of node-wise intensity functions and its adjacent nodes, with the node-wise intensity functions, adjusted for the weights used. That is, the Moran's $I$ statistic provides information on the correlation between the node-wise intensity functions, and the neighboring intensity function values.

Another concept of node-wise auto-correlation along spatial networks which uses the sum of squared differences between pairs of node-wise intensity functions as its measure of covariation is provided by Geary's $C$ statistic, 
\begin{equation}\label{eq:GearyC}
C = \frac{(n-1)}{2\sum_{i=1}^{n}\sum_{j=1}^{n}w_{ij}}
\frac{\sum_{i=1}^{n}\sum_{j=1}^{n}w_{ij}(\lambda(v_i)-\lambda(v_j))^2}{\sum_{i=1}^{n}(\lambda(v_i) - \m)^2}
\end{equation}
and, additionally, by Getis' and Orb's $G$ statistics,
\[
G = \frac{\sum_j w_{ij}\lambda(v_j)-\sum_{j}w_{ij}\m}{n\sum_jw_{ij}^2-(\sum_j w_{ij})^2/(n-1)^\frac{1}{2}}.
\]

Apart from these global statistics, we define a local counterpart of \eqref{eq:MoranI} as \[
I_i = \frac{(\lambda(v_i)-\m)}{{\sum_{j=1}^{n}(\lambda(v_j)-\m)^2}/(n-1)}{\sum_{j=1}^{n}w_{ij}(\lambda(v_j)-\m)}
\] 
and, following the ideas of \cite{Anselin1995}, a local version of \eqref{eq:GearyC} as 
\[
C_i=\frac{1}{\sum_{i=1}^n\lambda(v_i)^2/n}\frac{\sum_{i=1}^{n}\sum_{j=1}^{n}(\lambda(v_i)-\lambda(v_j))^2}{\sum_{i=1}^n\lambda(v_i)^2}
\]

\subsection{Local indicators of spatial network associations of type 2}

We now consider LISNA functions of type 2 which can be understood as a  generalization of Section \ref{sec:2ndorder} to second-order characteristics which describe variations in second-order network intensity functions for cross-hierarchical pairs of network entities. In general, any such cross-hierarchical pair consists of one edge interval and one subset of diverse edge intervals such as neighbors, parents or paths. By this, different from the previous section, LISNA functions of type 2 are not restricted to node-wise characteristics. But, while LISNA functions of type 1 are based on averaged and non-averaged node-wise first-order intensity functions, the present LISNA functions are related to the second-order properties of point patterns of spatial networks and only allow for non-averaged intensity functions. 

In general, this section only considers  LISNA functions for two different second-order properties:  the LISNA function with respect to (a) second-order non-average intensity functions and (b) auto-covariance functions.  

For (a), a generalization of Section \ref{sec:2ndorder} to cross-hierarchical terms yields to
\[
\lambda(\nach(v_i),s_{e_j})=\lim_{|d\flat(v_i),ds_{e_j}|\rightarrow 0}\left\{\frac{\mathds{E}\left[N(d\flat(v_i)),N(ds_{e_j})\right]}{|d\flat(v_i)\times ds_{e_j}|}\right\}.
\]
 Similarly, for (b) we have
\[
\gamma(\nach(v_i), s_{e_j})=\lambda(\nach(v_i), s_{e_j})-\lambda(\nach(v_i))\lambda(s_{e_j}).
\] 

A detailed list of all possible cross-hierarchical LISNA configurations for spatial networks is given in Table \ref{Tab:LINA1}. 

\begin{table}
\caption{Configurations of LINSA functions of type 2 for  different types of spatial networks. \label{Tab:LINA1}} 
\begin{center}
  \begin{tabular}{rrrr}
 type of network & $2^{nd}$ order intensity &  auto-covariance & Counting measures \\
 \hline   
undirected & $\lambda(s_{e_i},\nach(v_i))$ & $\gamma(s_{e_i}, s_{e_j})$  & $N(ds_{e_i}),  N(d\flat(v_i))$  \\ 
undirected & $\lambda(s_{e_i}, \pi^*_{ij})$& $\gamma(s_{e_i}, \pi^*_{ij})$ & $N(ds_{e_i}^{in}), N(d\wp_{ij})$\\
directed & $\lambda(s_{e_i}^{in}, \pa(v_i))$ & $\gamma(s_{e_i}^{in},  \pa(v_i))$ & $N(ds_{e_i}^{in}),  N(d\flat^{in}(v_i))$\\
directed & $\lambda(s_{e_i}^{out}, \pa(v_i))$ & $\gamma(s_{e_i}^{out},  \pa(v_i))$ & $N(ds_{e_i}^{out}),  N(d\flat^{in}(v_i))$\\
directed & $\lambda(s_{e_i}^{in}, \child(v_i))$& $\gamma(s_{e_i}^{in}, \child(v_i))$ & $N(ds_{e_i}^{in}), N(d\flat^{out}(v_i))$ \\
directed & $\lambda(s_{e_i}^{out}, \child(v_i))$& $\gamma(s_{e_i}^{out}, \child(v_i))$ & $N(ds_{e_i}^{out}), N(d\flat^{out}(v_i))$ \\
directed & $\lambda(s_{e_i}^{in}, \family(v_i))$& $\gamma(s_{e_i}^{in}, \family(v_i))$ & $N(ds_{e_i}^{in}), N(d\flat^{fam}(v_i))$ \\
directed & $\lambda(s_{e_i}^{out}, \family(v_i))$& $\gamma(s_{e_i}^{out}, \family(v_i))$ & $N(ds_{e_i}^{out}), N(d\flat^{fam}(v_i))$ \\
directed & $\lambda(s_{e_i}^{in}, \dpath^*_{ij})$& $\gamma(s_{e_i}^{in}, \dpath^*_{ij})$ & $N(ds_{e_i}^{in}), N(d\dwp_{ij}))$\\
directed & $\lambda(s_{e_i}^{out}, \dpath^*_{ij})$& $\gamma(s_{e_i}^{out}, \dpath^*_{ij})$ & $N(ds_{e_i}^{out}), N(d\dwp_{ij}))$\\
directed & $\lambda(s_{e_i}^{in}, \anch(v_i))$& $\gamma(s_{e_i}^{in}, \anch(v_i))$ & $N(ds_{e_i}^{in}), N(d\wp^{-i}_{ji}))$ \\
directed & $\lambda(s_{e_i}^{out}, \anch(v_i))$& $\gamma(s_{e_i}^{out}, \anch(v_i))$ & $N(ds_{e_i}^{out}), N(d\wp^{-i}_{ji}))$ \\
directed & $\lambda(s_{e_i}^{in}, \dech(v_i))$& $\gamma(s_{e_i}^{in}, \dech(v_i))$ & $N(ds_{e_i}^{in}), N(d\wp^{-i}_{ij})$ \\
directed & $\lambda(s_{e_i}^{out}, \dech(v_i))$& $\gamma(s_{e_i}^{out}, \dech(v_i))$ & $N(ds_{e_i}^{out}), N(d\wp^{-i}_{ij})$ \\
partially directed & $\lambda(s_{e_i}^{in}\,\nach(v_i))$ & $\gamma(s_{e_i}^{in}, s_{e_j})$  & $N(ds_{e_i}^{in}),  N(d\flat(v_i)))$  \\ 
partially directed & $\lambda(s_{e_i}^{out},\nach(v_i))$ & $\gamma(s_{e_i}^{out}, s_{e_j})$  & $N(ds_{e_i}^{out}),  N(d\flat(v_i))$  \\
partially directed & $\lambda(s_{e_i}^{in},\nach(v_i))$ & $\gamma(s_{e_i}^{in}, s_{e_j})$  & $N(ds_{e_i})^{in},  N(d\flat(v_i))$  \\
partially directed & $\lambda(s_{e_i}^{out}, \pi^*_{ij})$& $\gamma(s_{e_i}^{out}, \pi^*_{ij})$ & $N(ds_{e_i}^{out}), N(d\wp_{ij})$\\
partially directed & $\lambda(s_{e_i}, \pa(v_i))$ & $\gamma(s_{e_i},  \pa(v_i))$ & $N(ds_{e_i}),  N(d\flat^{in}(v_i))$\\
partially directed & $\lambda(s_{e_i}, \child(v_i))$& $\gamma(s_{e_i}, \child(v_i))$ & $N(ds_{e_i}), N(d\flat^{out}(v_i))$ \\
partially directed & $\lambda(s_{e_i}, \family(v_i))$& $\gamma(s_{e_i}, \family(v_i))$ & $N(ds_{e_i}), N(d\flat^{fam}(v_i))$ \\
partially directed & $\lambda(s_{e_i}, \dpath^*_{ij})$& $\gamma(s_{e_i}, \dpath^*_{ij})$ & $N(ds_{e_i}), N(d\dwp_{ij})$\\
partially directed & $\lambda(s_{e_i}, \anch(v_i))$& $\gamma(s_{e_i}, \anch(v_i))$ & $N(ds_{e_i}), N(d\wp^{-i}_{ji})$ \\
partially directed & $\lambda(s_{e_i}, \dech(v_i))$& $\gamma(s_{e_i}, \dech(v_i))$ & $N(ds_{e_i}), N(d\wp^{-i}_{ij})$ \\
  \end{tabular}
\end{center}
\end{table}

\section{Application: Urban disturbances-related data}\label{sec:apl}

This section covers applications of LISNA type 1 functions to spatial network data on locations of phone calls on neighbor and community disturbance recorded by local police authorities in the City of Castell{\'o}n.

\subsection{Data and network}
Our study is based on event data recorded along the traffic network of the City of Castell{\'o}n (Spain) for which we defined $1611$ segmenting units. Each segmenting unit is treated as endpoint of an edge interval such that each edge intervals is spanned between a pair of vertices, namely between two distinct segmenting units.  By this, we obtain a spatial network with a mean number of adjacent nodes of 3.14.  Next, we augmented each vertex with precise co-ordinates and computed the length of the edge interval as the squared geodesic distance between its geo-coded endpoints.  
For our analysis, we considered a geo-referenced subsample of N= $9790$ call-in events provided by the local officials of the City of Castell{\'o}n (Spain). Classification as neighbor and community disturbance has been performed prior to our analysis by police officials. The phone calls have been received at local police stations or transferred by 112 emergency services to local police call centers and geo-referenced indirectly by the provider based on precise address information.  Using this geo-information, we considered an event to belong to a distinct edge interval if the co-ordinates fell in-between the geo-coded endpoints of an edge. 
Adopting the network intensity function formalism to the resulting spatial network pattern, we computed edgewise mean intensity functions for all edge intervals contained in traffic network and calculated the node-wise first-order mean intensity function for neighboring nodes. By this, we obtained average node-wise intensity values for $614$ segmenting units which have been treated as input for the LISNA type 1 functions.

\subsection{Global and local associations for neighbor and community disturbances}
\begin{figure}[h!]
\begin{center}
\includegraphics[width=3in]{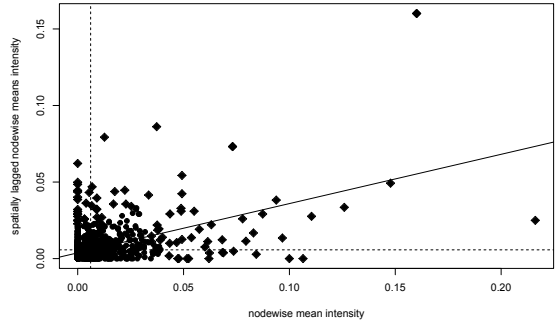}
\caption{Moran's $I$ scatterplot for the Castell{\'o}n network.}
\label{fig:MoranScatter}
\end{center}
\end{figure}

To evaluate the associations among node-wise first-order mean intensity values along the network, we first computed the Moran $I$ and the Geary's $C$ statistic. For Moran's $I$ we obtained a value of 0.32 and for Geary's $C$ a value of 0.58 which both indicate a positive auto-correlation in the distribution of node-wise mean intensity functions along the network, although it is not particularly strong. 
Besides the numerical characteristics, we also computed Moran's $I$ scatterplot \citep{Anselin1996} which is shown in Figure \ref{fig:MoranScatter}. This plot compares the node-wise first-order mean intensity function of each segmenting unit with the average value of its $1^{st}$ order neighboring nodes. The Moran $I$ statistics is depicted as the slope in the scatterplot with the neighboring node-wise   intensity value on the vertical and the node-wise   intensity function on the horizontal axis. Inspecting Figure, we found that almost all points in the scatterplot are placed in the upper right quadrant which confirms our findings of positive auto-correlation, where the slope of the regression lines indicates a moderate Moran $I$ statistic. 

\begin{figure}[h!]
\begin{center}
\subfloat[]{\label{fig:MoranCorrelogram}
\includegraphics[width=3in]{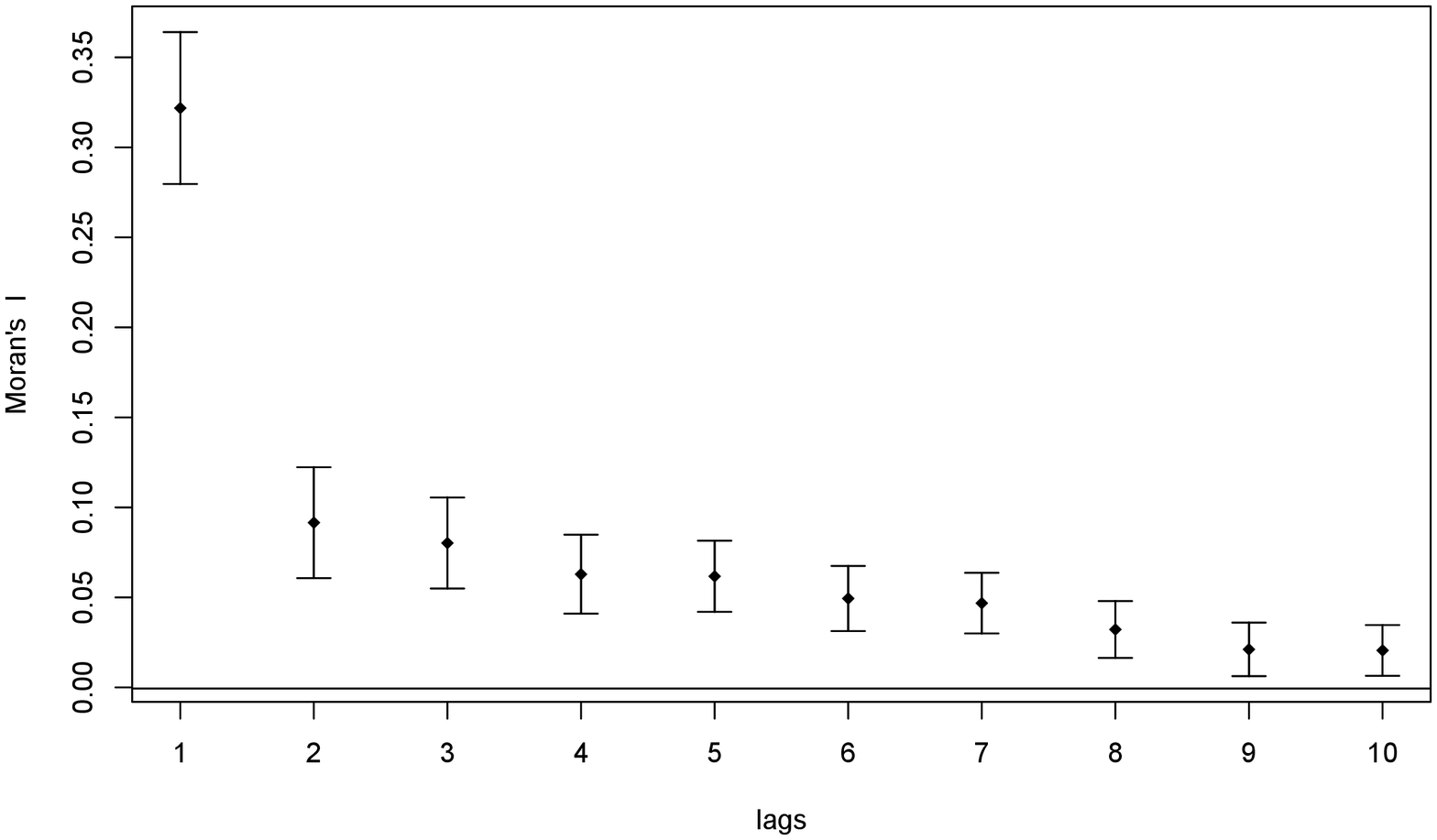}}
\subfloat[]{\label{Fig:PvalM}
\includegraphics[width=3in]{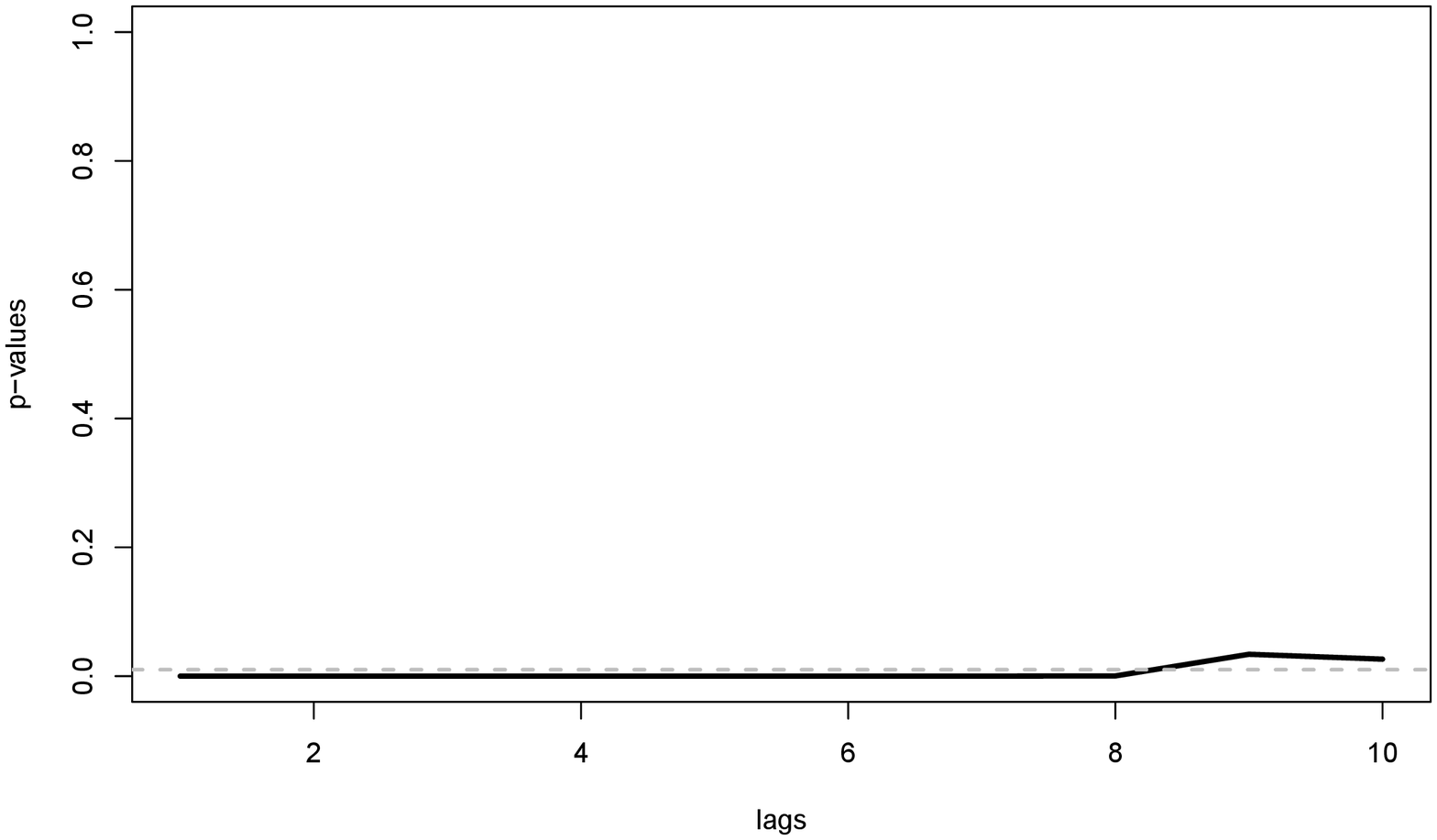}}\\
\subfloat[]{\label{fig:GearyCorrelogram}
\includegraphics[width=3in]{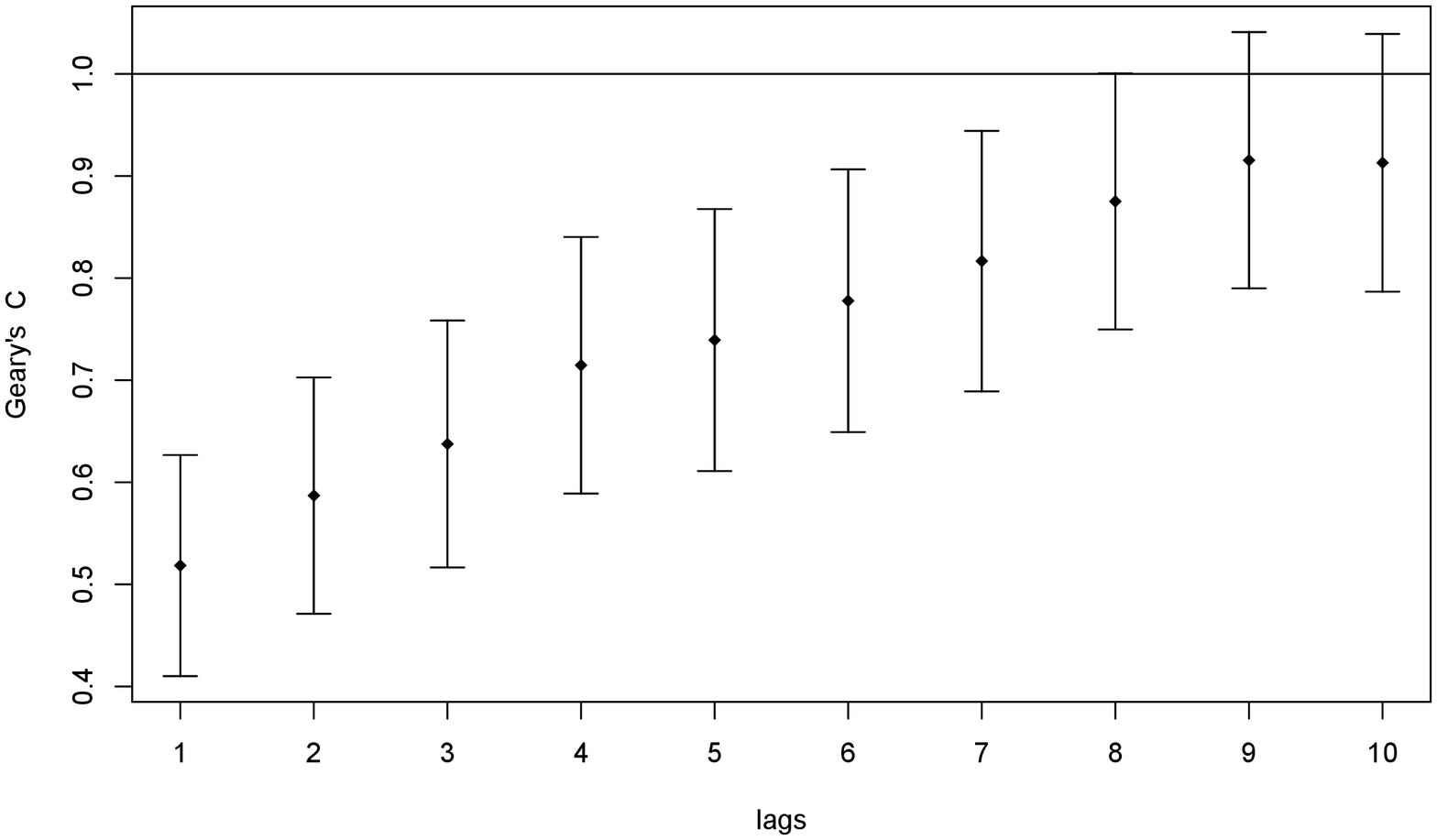}}
\subfloat[]{\label{Fig:PvalC}
\includegraphics[width=3in]{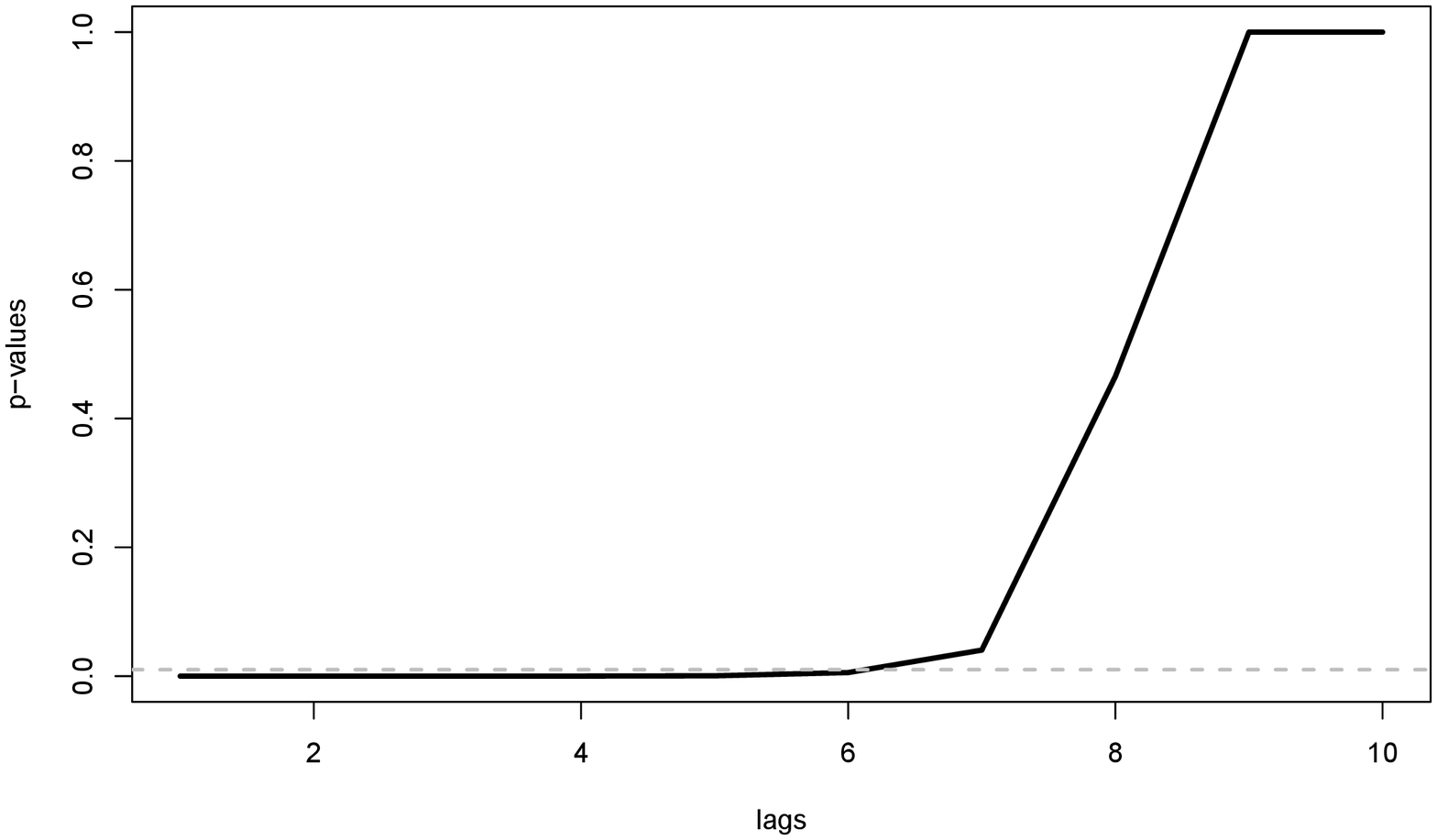}}
\caption{Correlograms and Bonferoni adjusted $p$-values for the Castell{\'o}n network: (a) Moran's $I$, (b) $p$-values of Moran's $I$ statistics (solid line) and $\alpha$-level of $0.99$ (dashed line), (c) Geary's $C$ and (d) $p$-values of Geary's $C$ statistics and $\alpha$-level of $0.99$ (dashed line).  \label{Fig:Corrplots}}
\end{center}
\end{figure}

To examine the order of spatial auto-correlation along the network structure, we additionally computed correlograms and Bonferoni adjusted $p$-values for the Moran's $I$ and Geary's $C$ statistic. The results are shown in Figure \ref{Fig:Corrplots}. Both correlograms show a consistent trend for Moran's $I$ and Geary's $C$ statistic indicating the presence of a positive auto-correlation among node-wise mean intensity functions along the network.  Looking at the $p$-values, we found a positive association among neighboring vertices up to order $6$ for Geary's $C$ statistic and up to order $8$ for Moran's $I$ statistic. 

\begin{figure}[h!]
\begin{center}
\includegraphics[width=3in]{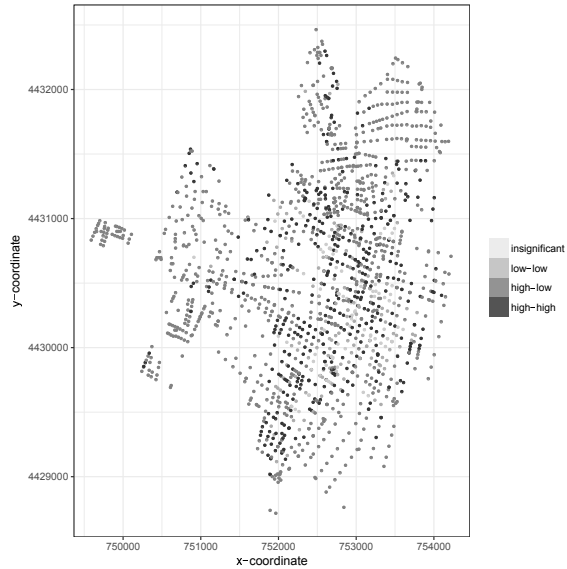}
\caption{Local  Moran's $I$ for the Castell{\'o}n network.}
\label{Fig:localI}
\end{center}
\end{figure}

 To further investigate the spatial auto-correlation among the node-wise mean intensity functions, we computed different local measures of auto-correlation. For the local Moran's $I$ statistic, as displayed in Figure \ref{Fig:localI}, we found high-low associations among the node-wise mean intensity functions of neighboring vertices in the upper- and lower-right areas as well as the left boarders  of the Castell{\'o}n traffic network. At the same time, high-high associations which reflect hotspots of node-wise mean call-in intensities occurred most frequently on a vertical axis along the central area of the traffic network. These findings express a severe clustering of neighbor and community disturbance call-ins along the downtown areas of Castell{\'o}n. 
Local Moran's $I$  for the Castell{\'o}n traffic network is shown in Figure 

\begin{figure}[h!]
\begin{center}
\includegraphics[width=3in]{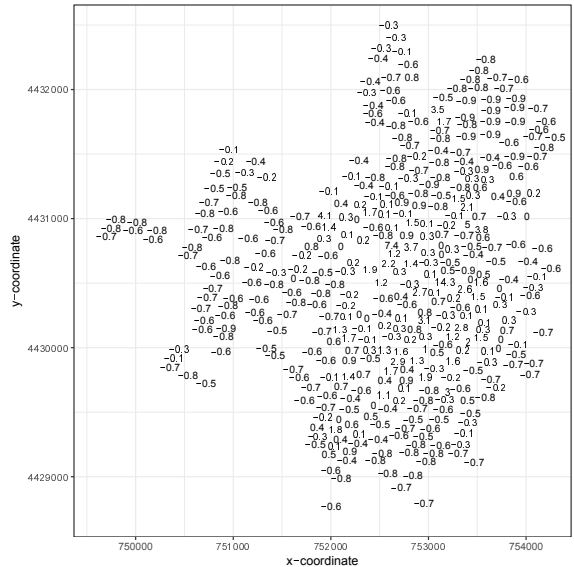}
\caption{Local Getis $G$ for the Castell{\'o}n network.}
\label{Fig:LocalG}
\end{center}
\end{figure}

Apart from the local Moran $I$ statistic, we concerned the local Getis' and Orb's $G$ statistic.  The results of the local Getis' and Orb's $G$ statistic are shown in Figure \ref{Fig:LocalG}. Different from Moran's $I$ or Geary's $C$ statistic, this local statistic also differentiates between high-high and low-low correlations which are treated as positive auto-correlation by Moran's $I$ statistic. Inspecting this Figure, we found high values located in the center whereas moderate low values occurred in the outlying areas of the Castell{\'o}n traffic network. These findings indicate a strong spatial agglomeration of neighbor and community disturbances such that all perturbations appeared within the central areas of Castell{\'o}n. One possible explanation for this local agglomeration of public disturbances can be seen in the denseness of the traffic network and the high population density in the city center.  

\section{Conclusion}\label{sec:conclusion}

This article has concerned the second-order analysis of structured point patterns over networks by means of network intensity functions and proposed node-wise and cross-hierarchical types of local indicator of network association functions. We believe that the presented methodology is immediately useful in the following sense, and could stimulate a rich body of future research and new directions in the analysis of point patterns and event-driven data recorded along planar networks. 

Having point data over planar line structures under study, one commonly faces heterogeneous rather than homogeneous characteristics along the network. The expected number of events is strongly associated with the specificity and geometrical complexity of the network and might be effected by the shape, the length and the characteristics of individual lines. Defining edge intervals to be the core elements, network intensity functions resolve any such methodological challenges and allows to explore the first- and higher-order characteristics of the point patterns under control of the network specificity. The proposed global and local network intensity functions provide information on interactions within and between different hierarchical levels contained in the network.

\bibliographystyle{Chicago}
\bibliography{lisa}
\end{document}